\documentclass[preprint,aps,prd,showpacs,floatfix]{revtex4-1}
\usepackage{times}
\usepackage {euscript}
\usepackage{amsthm}
\usepackage{graphicx}
\usepackage{amsmath}
\usepackage{amssymb}
\usepackage{amsfonts}
\usepackage[english]{babel}
\usepackage{epstopdf}
\usepackage{multirow,makecell,array}
\usepackage{mathtools}
\usepackage{epstopdf}
\begin{document}
\thispagestyle{empty}

    \title{Electron-positron pair production in external electric fields\\ varying both in space and time}

\author{I.~A.~Aleksandrov$^{1, 2}$}
\email[]{i.aleksandrov@spbu.ru}
\author{G.~Plunien$^{3}$} \author{V.~M.~Shabaev$^{1}$}
\affiliation{$^1$~Department of Physics, St. Petersburg State University, 7/9 Universitetskaya nab., Saint Petersburg 199034, Russia\\$^2$~ITMO University, Kronverkskii ave 49, Saint Petersburg 197101, Russia\\$^3$~Institut f\"ur Theoretische Physik, TU Dresden, Mommsenstrasse 13, Dresden, D-01062, Germany
\vspace{10mm}
}

\begin{abstract}
The Schwinger mechanism of electron-positron pair production in the presence of strong external electric fields is analyzed numerically for the case of one- and two-dimensional field configurations where the external field depends both on time and one spatial coordinate. In order to provide this analysis, a new efficient numerical approach is developed. The number of particles created is obtained numerically and also compared with the analytical results for several exactly solvable one-dimensional backgrounds. For the case of two-dimensional field configurations the effects of the spatial finiteness are examined, which confirms their importance and helps us to attest our approach further. The corresponding calculations are also performed for several more interesting and nontrivial combinations of temporal and spatial inhomogeneities. Finally, we discuss the case of a spatially periodic external field when the approach is particularly productive. The method employed is described in detail.
\end{abstract}


\maketitle
\section{Introduction}
\label{sec:intro}
The phenomenon of particle creation from vacuum in the presence of space-time-varying strong external fields has been a subject of a great amount of studies~\cite{klein2, sauter_pairs_1931, sauter_klein_1931, gav_git_x_electric_2016, schwinger_1951, fradkin_gitman_shvartsman, gmm_1994, greiner_qed_sf, nikishov_1969, narozhny_1970, bagrov_1975, gav_git_prd_1996, gelis_2015, gav_git_lconstant_2016, adorno_phys_scr_2015, adorno_2015, adorno_2016, euler_heisenberg, ruf_prl_2009, jiang_pra_2011, jiang_pra_2012, hebenstreit_prl_2011, hebenstreit_prd_2010, hebenstreit_prd_2011, linder_prd_2015, dumlu_prd_2016, schneider_2016, burke_1997, dunne_prd_2005, schutzhold_prl_2008, gies_prl_2016, woellert_2015, dunne_prd_2009}. Such a process arises due to the instability of the physical vacuum and can be rigorously described within the framework of quantum electrodynamics (QED). From relativistic quantum mechanics it follows that the characteristic critical field strength is $E_\text{c} = m^2 c^3/(|e| \hbar) \approx 1.3 \times 10^{16}$ V/cm ($m$ is the electron mass)~\cite{sauter_pairs_1931} and, therefore, it is very difficult to achieve such strong fields experimentally (so far the pair creation process has been observed only in the perturbative multiphoton regime~\cite{burke_1997}). Nevertheless, from a theoretical point of view both to formulate a stringent theory of QED in strong external backgrounds and to provide numerical estimations of the corresponding effects are of fundamental importance. The latter task is the focus of the present investigation.

\indent The first theoretical analyses of the problem were related to the Klein paradox described in Refs.~\cite{klein2, sauter_pairs_1931, sauter_klein_1931} where a static potential step was considered (see also Refs.~\cite{gav_git_x_electric_2016, gav_git_lconstant_2016} and references therein). In Ref.~\cite{schwinger_1951} the vacuum-vacuum transition probability was determined for the case of a constant and uniform electric field by means of the effective Lagrangian formalism (see also Ref.~\cite{euler_heisenberg}). For the case of time-dependent fields a general theoretical approach based on the quantization of charged fields within the Furry picture can be found in Ref.~\cite{fradkin_gitman_shvartsman} (see also Refs.~\cite{gmm_1994, nikishov_1969, narozhny_1970, bagrov_1975, gav_git_prd_1996, gelis_2015, adorno_phys_scr_2015, adorno_2015, adorno_2016}). Analytical expressions for the scattering probabilities and mean numbers of particles created have been derived only for very few one-dimensional (1D) configurations of the external field (see, e.~g., Ref.~\cite{adorno_2015}) where the field depends only on time or only on one spatial coordinate. However, in order to study more general scenarios one has to consider higher-dimensional cases for which the corresponding calculation procedures should be designed. During the last few years a number of attempts have been made to examine the Schwinger effect in space-time-dependent configurations of external fields by means of various techniques~\cite{ruf_prl_2009, jiang_pra_2011, jiang_pra_2012, hebenstreit_prl_2011, hebenstreit_prd_2010, hebenstreit_prd_2011, linder_prd_2015, dumlu_prd_2016, schneider_2016}. In Refs.~\cite{ruf_prl_2009, jiang_pra_2011, jiang_pra_2012} the Dirac equation, including the interaction with an external field, was solved numerically on a spatial grid. The corresponding solutions contain all the information needed to obtain observable quantities (this will be also discussed in Section~\ref{sec:theory}). In Ref.~\cite{hebenstreit_prl_2011} the Dirac-Heisenberg-Wigner formalism~\cite{hebenstreit_prd_2010, hebenstreit_prd_2011} was employed to study a simple two-dimensional (2D) background. In Refs.~\cite{linder_prd_2015, dumlu_prd_2016, schneider_2016} the world-line method~\cite{dunne_prd_2005} was used to investigate several other types of external electric fields. Nevertheless, our knowledge about 2D models is still very limited and, therefore, it is strongly desirable to develop new independent techniques for the corresponding analysis. In this paper we present an accurate and efficient numerical approach that allows us to investigate space- and time-dependent backgrounds nonperturbatively. The method is based on solving the Dirac equation in the momentum representation. We provide an analysis of several 2D types of the external field depending on both time and one spatial coordinate. The results are compared with those for the 1D case and the method is attested in the corresponding limits. Besides, our approach was applied to the field configurations considered in Ref.~\cite{jiang_pra_2012} where it was shown that a special combination of space-time-dependent pulses may significantly increase the pair-production rate. Our calculations reproduce the previous findings. Finally, it is demonstrated that spatially periodic external fields can be treated much more efficiently. All the examples indicate that the method presented in this paper allows one to study a broad class of different external backgrounds with multidimensional inhomogeneities and, thus, provides a deeper understanding of the pair-production phenomenon beyond the 1D case. Furthermore, it can be subsequently extended to a variety of other problems.

\indent We employ relativistic units ($\hbar = 1$, $c = 1$) and the Heaviside charge unit ($\alpha = e^2/4\pi$) throughout the paper and assume the electron to have the charge $e = -|e|$ and mass $m = 1$. Accordingly, the critical electric field strength is given by $|e| E_\text{c} = 1~\text{[ru]}$ and the unit of length is the ``reduced'' electron Compton wavelength $\lambdabar = \hbar/mc = 1~\text{[ru]} \approx 3.9 \times 10^{-11}$ cm. One relativistic unit of time is $\hbar/mc^2 \approx 1.3 \times 10^{-21}$~s.

\indent The outline of this paper is as follows. In Section~\ref{sec:theory} we present a brief description of the general approach employed throughout the paper. In Section~\ref{sec:one_dim} we discuss a simple way to obtain the mean numbers of particles for arbitrary 1D electric fields (i.~e., uniform time-dependent fields). In the main part of the paper (Section~\ref{sec:two_dim}) we describe an efficient calculation scheme for the case of space-time-varying external fields and present the results of our numerical calculations. In Section~\ref{sec:bose} we briefly discuss the case of scalar particles. Finally, in Section~\ref{sec:discussion}, we provide a summary of the study.
\section{General approach}
\label{sec:theory}
First, we will briefly describe how the interaction with time-dependent external fields can be taken into account exactly, i.~e., within the Furry picture. This approach is formulated in detail in Ref.~\cite{fradkin_gitman_shvartsman} (for the case of static space-dependent electric fields the quantization procedure is different~\cite{gav_git_x_electric_2016}). The Dirac equation in the presence of an external field contains the minimal coupling terms:
\begin{equation}
\big ( \gamma^\mu \big [i \partial_\mu - e A_\mu (t, \boldsymbol{x}) \big ] - m \big) \Psi (t, \boldsymbol{x}) = 0, \quad \partial_\mu = (\partial_t, \boldsymbol{\nabla}),~A_\mu = (A_0, -\boldsymbol{A}).
\label{eq:dirac_general}
\end{equation}
We consider $(3+1)$-dimensional QED, where $\boldsymbol{x}$ denotes a three-dimensional spatial vector and $x^0=t$ is the temporal component of the four-vector $x$. We invoke the temporal gauge $A_0 (x) = 0$, so Eq.~(\ref{eq:dirac_general}) can be rewritten as
\begin{equation}
\big [ i\partial_0 - \hat{h}(x) \big] \Psi (x) = 0,~~\text{where}~~\hat{h} (x) = \boldsymbol{\alpha} \cdot \big[-i\boldsymbol{\nabla} - e \boldsymbol{A} (x)\big ] + \beta m.
\label{eq:dirac_h}
\end{equation}
The time-dependent external field is assumed to be switched on at $t = t_\text{in}$ and switched off at $t = t_\text{out}$. We introduce in and out orthonormal and complete sets of solutions of Eq.~(\ref{eq:dirac_general}) \big($\{ {}_\zeta \Psi_n (t, \boldsymbol{x})\}$ and $\{ {}^\zeta \Psi_n (t, \boldsymbol{x})\}$, $\zeta = \pm$, respectively\big) which obey
\begin{eqnarray}
&&{}_\zeta \Psi_n (t_\text{in}, \boldsymbol{x}) = {}_\zeta \Psi^{(0)}_n (\boldsymbol{x}),~{}^\zeta \Psi_n (t_\text{out}, \boldsymbol{x}) = {}^\zeta \Psi^{(0)}_n (\boldsymbol{x}),\label{eq:psi_in_out}\\
&&({}_\zeta \Psi_n, \, {}_\kappa \Psi_m) = ({}^\zeta \Psi_n, \, {}^\kappa \Psi_m) = \delta_{\zeta \kappa} \delta_{nm}\quad \text{for all}~t,  \label{eq:psi_orthonormality} \\
&&\sum_{n, \zeta} {}_\zeta \Psi_n (t, \boldsymbol{x}) \, {}_\zeta \Psi^\dagger_n (t, \boldsymbol{x}') = \sum_{n, \zeta} {}^\zeta \Psi_n (t, \boldsymbol{x}) \, {}^\zeta \Psi^\dagger_n (t, \boldsymbol{x}') = I\delta (\boldsymbol{x} - \boldsymbol{x}') \label{eq:psi_completeness},
\end{eqnarray}
where the inner product is given by $(\Psi_1, \, \Psi_2) = \int \Psi^\dagger_1 \Psi_2 \mathrm{d}^3\boldsymbol{x}$ and $I$ is the $4\times 4$ identity matrix. The functions ${}_\zeta \Psi^{(0)}_n (\boldsymbol{x})$ and ${}^\zeta \Psi^{(0)}_n (\boldsymbol{x})$ are the eigenfunctions of the Dirac Hamiltonian $\hat{h}(x)$ considered at times $t_\text{in}$ and $t_\text{out}$, respectively. They also compose orthonormal and complete sets and the sign $\zeta$ denotes the sign of the eigenvalues. The field operator can be expanded either into the basis of the in solutions, or into the out set. This means that there are two sets of the electron/positron creation and annihilation operators defined with respect to the two vacua (in and out). One can establish the following connection between the in set and out set of the creation and annihilation operators~\cite{fradkin_gitman_shvartsman}:
\begin{eqnarray}
&&\hat{b}_n (\text{out}) = \sum_m \hat{b}_m (\text{in}) G({}^+|{}_+)_{nm} + \sum_m \hat{d}^\dagger_m (\text{in}) G({}^+|{}_-)_{nm},\label{eq:b_out}\\
&&\hat{d}^\dagger_n (\text{out}) = \sum_m \hat{b}_m (\text{in}) G({}^-|{}_+)_{nm} + \sum_m \hat{d}^\dagger_m (\text{in}) G({}^-|{}_-)_{nm},\label{eq:d_out}\\
&&\hat{b}_n (\text{in}) = \sum_m \hat{b}_m (\text{out}) G({}_+|{}^+)_{nm} + \sum_m \hat{d}^\dagger_m (\text{out}) G({}_+|{}^-)_{nm},\label{eq:b_in}\\
&&\hat{d}^\dagger_n (\text{in}) = \sum_m \hat{b}_m (\text{out}) G({}_-|{}^+)_{nm} + \sum_m \hat{d}^\dagger_m (\text{out}) G({}_-|{}^-)_{nm},\label{eq:d_in}
\end{eqnarray}
where the coefficients can be expressed as the inner products of the in and out solutions:
\begin{eqnarray}
&&G({}^\zeta|{}_\kappa)_{nm} = ({}^\zeta \Psi_n,~{}_\kappa \Psi_m), \label{eq:G_inner_product_1}\\
&&G({}_\zeta|{}^\kappa)_{nm} = ({}_\zeta \Psi_n,~{}^\kappa \Psi_m). \label{eq:G_inner_product_2}
\end{eqnarray}
Note, that these inner products are time independent since the operator $\hat{h}(x)$ is symmetric for all values of $t$. These $G$ matrices contain all the information about the scattering probabilities ($\zeta = \kappa$) and spectrum of particles created ($\zeta = -\kappa$). For instance, the number of electrons (positrons) produced with the given quantum numbers $m$ can be evaluated as follows:
\begin{eqnarray}
&&n^-_m = \langle 0,\text{in} | \hat{b}^\dagger_m (\text{out}) \hat{b}_m (\text{out}) |0,\text{in}\rangle = \sum_n G({}^+|{}_-)_{mn} G({}_-|{}^+)_{nm} = \{ G({}^+|{}_-) G({}_-|{}^+) \}_{mm}, \label{eq:num_el}\\
&&n^+_m = \langle 0,\text{in} | \hat{d}^\dagger_m (\text{out}) \hat{d}_m (\text{out}) |0,\text{in}\rangle = \sum_n G({}^-|{}_+)_{mn} G({}_+|{}^-)_{nm} = \{ G({}^-|{}_+) G({}_+|{}^-) \}_{mm}. \label{eq:num_pos}
\end{eqnarray}
The vacuum-vacuum transition probability is given by~\cite{fradkin_gitman_shvartsman}
\begin{equation}
p_v = |\langle 0,\text{out} | 0,\text{in}\rangle|^2 = |\mathrm{det}~G({}^-|{}_-)|^2 =  |\mathrm{det}~G({}^+|{}_+)|^2.
\label{eq:vacuum_vacuum}
\end{equation}
\indent In order to construct the in and out sets of solutions we will use the conventional substitution $\Psi = \big [ \gamma^\mu \big (i \partial_\mu - e A_\mu \big ) + m \big] \psi$ which leads to the following equation (see, e. g., Refs.~\cite{fradkin_gitman_shvartsman, gav_git_prd_1996}):
\begin{equation}
\big ( [i \partial - e A ]^2 - m^2 - \frac{ie}{2} \gamma^\mu \gamma^\nu F_{\mu \nu} \big ) \psi (t, \boldsymbol{x}) = \big ( [i \partial - e A]^2 - m^2 - \frac{e}{2} \sigma^{\mu \nu} F_{\mu \nu} \big ) \psi (t, \boldsymbol{x}) = 0,
\label{eq:dirac_squared}
\end{equation}
where $\sigma^{\mu \nu} = \frac{i}{2} [\gamma^\mu, \gamma^\nu]$ and $F_{\mu \nu} = \partial_\mu A_\nu - \partial_\nu A_\mu$. Accordingly, we can write $\frac{e}{2} \sigma^{\mu \nu} F_{\mu \nu} = ie\, \boldsymbol{\alpha} \cdot \boldsymbol{E} - e\, \boldsymbol{\Sigma} \cdot \boldsymbol{B}$. In the present paper, in and out solutions will be obtained numerically for certain space- and time-dependent configurations of the external field.
\section{One-dimensional case}\label{sec:one_dim}
First, we consider a spatially homogeneous field directed along the $x$ axis: $E_x = E(t)$. In the temporal gauge:
\begin{equation}
A_0 = 0, \quad A_y = A_z = 0,\quad A_x (t) = - \int \limits^t E(t') \mathrm{d} t'.
\label{eq:gauge_t}
\end{equation}
This allows one to rewrite Eq.~(\ref{eq:dirac_squared}) in the form
\begin{equation}
\big (\partial_t^2 - \Delta + 2 ie A_x (t) \partial_x + e^2 A_x^2 (t) + m^2 + ie \gamma^0 \gamma^1 E (t) \big ) \psi_n (t, \boldsymbol{x}) = 0.
\label{eq:dirac_t}
\end{equation}
The function $\psi_n (t, \boldsymbol{x})$ can be expressed as~\cite{fradkin_gitman_shvartsman}
\begin{equation}
\psi_n (t, \boldsymbol{x}) = \psi_{\boldsymbol{p}, s, r} (t, \boldsymbol{x}) = \mathrm{e}^{i\boldsymbol{p} \boldsymbol{x}} v_{s,r} \varphi_{\boldsymbol{p}, s, r} (t),
\label{eq:psi_t_spin}
\end{equation}
where $v_{s,r}$ ($s = \pm 1$, $r = \pm 1$) is a set of constant orthonormalized spinors that are the eigenvectors of the matrix $\gamma^0 \gamma^1 = \alpha^1$:
\begin{equation}
\gamma^0 \gamma^1 v_{s,r} = s v_{s,r},\quad v_{s,r}^\dagger v_{s',r'} = \delta_{s,s'} \delta_{r,r'}.
\label{eq:spinors}
\end{equation}
For the scalar function $\varphi_{\boldsymbol{p}, s, r} (t)$ Eq.~(\ref{eq:dirac_t}) reads
\begin{equation}
\big (\partial_t^2 + (p_x - e A_x (t))^2 + \pi_\perp^2 + iesE(t) \big ) \varphi_{\boldsymbol{p}, s, r} (t) = 0,\quad \pi_\perp^2 = \boldsymbol{p}_\perp^2 + m^2, \quad \boldsymbol{p} = (p_x, \boldsymbol{p}_\perp).
\label{eq:dirac_t_phi}
\end{equation}
\begin{sloppypar}
By solving this ordinary differential equation one can obtain the two sets $\{ {}_\pm \Psi_n (t, \boldsymbol{x})\}$ and $\{ {}^\pm \Psi_n (t, \boldsymbol{x})\}$ and calculate the matrix elements given by Eqs.~(\ref{eq:G_inner_product_1}) and (\ref{eq:G_inner_product_2}). In the case under consideration these matrices are diagonal: $G ({}^\kappa|{}_\zeta)_{mn} = \delta_{mn} \, g ({}^\kappa|{}_\zeta)_{n}$ or ${G ({}^\kappa|{}_\zeta)_{\boldsymbol{p}, r;\, \boldsymbol{p}^\prime, r^\prime} = \delta (\boldsymbol{p} - \boldsymbol{p}^\prime) \, \delta_{r,r^\prime} \, g ({}^\kappa|{}_\zeta)_{\boldsymbol{p}, r}}$. Therefore, one can evaluate, e.~g., the number density $n^-_{\boldsymbol{p},r}$ of particles (electrons) of momentum $\boldsymbol{p}$ via
\end{sloppypar}
\begin{equation}
 \frac{(2\pi)^3}{V} \, \frac{\mathrm{d} N_{\boldsymbol{p},r}}{\mathrm{d}^3 \boldsymbol{p}} = \frac{(2\pi)^3}{V} \, n^-_{\boldsymbol{p},r} = |g ({}_-|{}^+)_{\boldsymbol{p},r}|^2.
\label{eq:t_coeff2}
\end{equation}
Here the system is assumed to have a finite volume $V$ and, therefore, the common substitution $\delta (\boldsymbol{p} = \boldsymbol{0}) = V/(2\pi)^3$ can be used. If one does not distinguish states with different values of $r$, then the result should be multiplied by $2$: $n^-_{\boldsymbol{p}} = 2n^-_{\boldsymbol{p},r}$. The labels $n = (\boldsymbol{p}, \, r)$ of the functions ${}^\zeta \varphi_n$ and ${}_\zeta \varphi_n$ do not include $s$, since the solutions with different $s$ are dependent (see Ref.~\cite{gav_git_prd_1996}): we use $s = \pm 1$ for ${}_\pm \varphi_n$ and ${}^\mp \varphi_n$, respectively. One can explicitly verify that this prescription is in agreement with Eqs.~(\ref{eq:psi_orthonormality}) and~(\ref{eq:psi_completeness}). In fact, these functions do not depend on $r$ as Eq.~(\ref{eq:dirac_t_phi}) does not contain its value. It is also worth noting that $p_x$ is the $x$ component of the generalized momentum which is not gauge invariant ($p^\text{gen}_x = p_x$). The ``physical'' gauge-invariant momentum of the electron is given by $p^\text{inv}_x = p^\text{gen}_x - eA_x$ (in order to obtain the spectrum of positrons produced in terms of their ``physical'' momenta one should use the positron charge $-e = |e|$).

\indent The method was applied to several different configurations of the external field. For the analytically solvable cases discussed in Refs.~\cite{adorno_phys_scr_2015, adorno_2015, adorno_2016, gelis_2015} our calculations reproduce the exact values with perfect accuracy. It is possible to consider an arbitrary function $E(t)$ and, moreover, this technique can be easily modified in order to treat arbitrary static fields $E(x)$, which are inhomogeneous in one space direction~\cite{gav_git_x_electric_2016, gav_git_lconstant_2016}. For instance, for the case of the Sauter-like space-dependent field $E(x) = E/\!\cosh^2 (x/\alpha)$ our results are in excellent agreement with the analytical expressions from Ref.~\cite{gav_git_x_electric_2016}.

\indent Nevertheless, more realistic backgrounds contain both temporal and spatial inhomogeneities and, thus, the corresponding analysis becomes a very difficult task. An efficient numerical approach for the 2D case is described in the next section.
\section{Two-dimensional case}\label{sec:two_dim}
\subsection{Method description}\label{subsec:two_dim_method}
\begin{figure}[h]
\center{\includegraphics[width=0.7\linewidth]{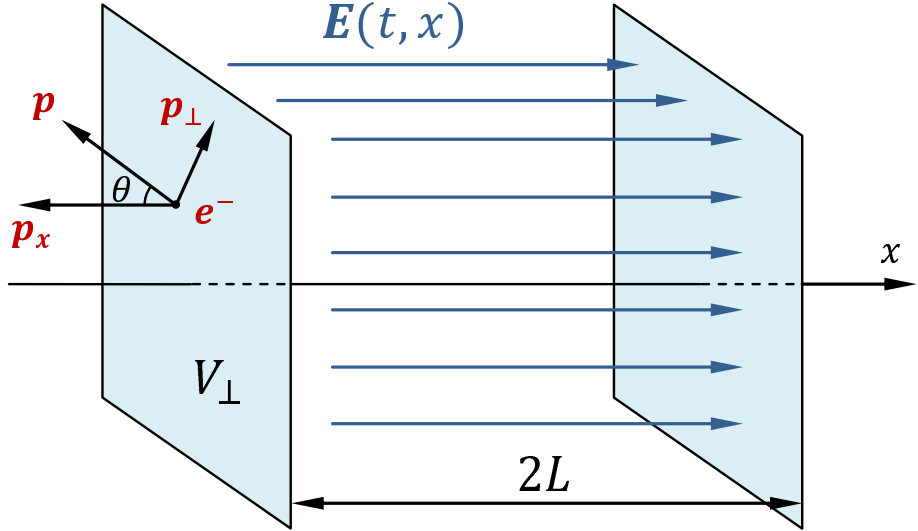}}
\caption{Illustration of the two-dimensional field configuration. The electric field is confined in the space layer, which has a width of $2L$, and is assumed to be switched on and off at the times $t_\text{in}$ and $t_\text{out}$, respectively. The electron created has a longitudinal component of its momentum $\boldsymbol{p}_x$ and a transverse component $\boldsymbol{p}_\perp$ which is orthogonal to the $x$ axis.}
\label{fig:capacitor}
\end{figure}

In this section we present a technique that can be used in order to obtain the quantities discussed for the case of electric fields $E_x = E (t, x)$ which depend on both the time $t$ and space coordinate $x$. We assume that the corresponding function $E(t, x)$ vanishes if $x \notin [-L, L]$ or $t \notin [t_\text{in}, t_\text{out}]$. The field configuration may be viewed as a capacitor with infinite plates (see Fig.~\ref{fig:capacitor}). The corresponding solution $\psi (t, \boldsymbol{x})$ of Eq.~(\ref{eq:dirac_squared}) can be represented as
\begin{equation}
\psi_{\boldsymbol{p}, s, r} (t, \boldsymbol{x}) = \mathrm{e}^{i\boldsymbol{p}_\perp \boldsymbol{x}_\perp} v_{s,r} \varphi_{\boldsymbol{p}, s, r} (t, x).
\label{eq:psi_xt_spin}
\end{equation}
The scalar function $\varphi_{\boldsymbol{p}, s ,r}$ now depends on both $t$ and $x$. We will omit the index $\boldsymbol{p}_\perp$, since the perpendicular component of the momentum $\boldsymbol{p}$ is conserved for such configurations (it can be treated as an effective mass $\pi_\perp^2 = m^2 + \boldsymbol{p}_\perp^2$). The value of $s$ will be chosen according to the rule mentioned in the previous section ($s = \pm 1$ for ${}_\pm \varphi$ and ${}^\mp \varphi$, respectively). The quantum number $r$ should be taken into account by multiplying the result by a factor of $2$: $n^{\pm}_{\boldsymbol{p}} = 2n^{\pm}_{\boldsymbol{p}, r}$. The function $\varphi (t, x)$ will carry only one index $p_x$ whose meaning will be discussed below. This function $\varphi_{p_x} (t, x)$ should be a solution of the following two-dimensional equation:
\begin{equation}
\big (\partial_t^2 - \partial_x^2 + 2ieA_x (t, x) \partial_x + ie \partial_x A_x (t, x) + e^2 A^2_x (t, x) + ies E(t, x) + \pi_\perp^2 \big ) \varphi_{p_x} (t, x) = 0.
\label{eq:dirac_xt_phi}
\end{equation}
Again the temporal gauge $A_0 (t, x) = 0$ is employed.

\indent The asymptotic behavior of the in and out solutions can be easily derived by solving the equation for asymptotic times $t \leq t_\text{in}$ and $t \geq t_\text{out}$, when the electromagnetic potential is a pure gauge function:
\begin{eqnarray}
{}_\pm \varphi_{p_x} (t, x) &=& {}_\pm C_{p_x} \mathrm{e}^{\mp(ip_0 (t - t_\text{in}) - ip_x x)}\quad \text{for}~t \leq t_\text{in}, \label{eq:phi_in} \\
{}^\pm \varphi_{p_x} (t, x) &=& {}^\pm C_{p_x} \mathrm{exp} \bigg ( ie \int \limits_0^x \! A_x (t_\text{out}, x^\prime) \mathrm{d} x^\prime \bigg )\mathrm{e}^{\mp(ip_0 (t - t_\text{out}) - ip_x x)}\quad \text{for}~t \geq t_\text{out}, \label{eq:phi_out}
\end{eqnarray}
where $p_0 = \sqrt{p_x^2 + \pi_\perp^2} > 0$. The values of $p_x$ now relate to the gauge-invariant momentum of the electron: $p^\text{inv}_x = p_x = p^\text{gen}_x - e A_x$. The normalization coefficients ${}_\pm C_{p_x}$ and ${}^\pm C_{p_x}$ can be found from Eqs.~(\ref{eq:psi_orthonormality}) and~(\ref{eq:psi_completeness}):
\begin{equation}
{}_\pm C_{p_x} = {}^\mp C_{p_x} = (2\pi )^{-3/2} \, \big [2p_0 (p_0 \mp p_x)\big ]^{-1/2}.
\label{eq:C_in_out}
\end{equation}
The main idea of the method is to solve  the problem in time-momentum space. The Fourier transform of a given out solution represents it as a combination of the functions $\mathrm{e}^{-ikx}$ which are essentially the in solutions~(\ref{eq:phi_in}) in the region $t \leq t_\text{in}$. Therefore, in momentum space one can propagate the $G$ matrix itself without using Eqs.~(\ref{eq:G_inner_product_1}) and~(\ref{eq:G_inner_product_2}). Besides, the corresponding Fourier transforms can be easily treated inside a finite $k$ box while the functions~(\ref{eq:phi_in}) and~(\ref{eq:phi_out}) have an infinite support. We express the functions ${}_\pm \varphi_{p_x}$ and ${}^\pm \varphi_{p_x}$ as
\begin{eqnarray}
{}_\pm \varphi_{p_x} (t, x) &=& \int \limits_{-\infty}^\infty \! \mathrm{d}k \, \mathrm{e}^{-ikx}\, {}_\pm f_{p_x} (t, k), \label{eq:fourier_in} \\
{}^\pm \varphi_{p_x} (t, x) &=& \int \limits_{-\infty}^\infty \! \mathrm{d}k \, \mathrm{e}^{-ikx}\, {}^\pm f_{p_x} (t, k). \label{eq:fourier_out}
\end{eqnarray}
The behavior of the spatial Fourier transform ${}_\pm f_{p_x}$ for asymptotic times $t \leq t_\text{in}$ reads
\begin{equation}
{}_\pm f_{p_x} (t, k) = {}_\pm C_{p_x} \mathrm{e}^{\mp ip_0 (t - t_\text{in})} \, \delta (k \pm p_x).
\label{eq:f_in}
\end{equation}
The functions ${}^\pm f_{p_x}$ can be found explicitly for the particular forms of $A_x (t, x)$. In time-momentum space Eq.~(\ref{eq:dirac_xt_phi}) for the function $f_{p_x}$ takes the form of an integro-differential equation:
\begin{equation}
\partial_t^2 f_{p_x} (t, k) + g(k) f_{p_x} (t, k) + \int \limits_{-\infty}^\infty \! \mathrm{d}q L(t, k, q) f_{p_x} (t, q) = 0,
\label{eq:dirac_xt_phi_momentum}
\end{equation}
where
\begin{eqnarray}
g(k) &=& \pi_\perp^2 + k^2, \label{eq:dirac_int_g} \\
L (t, k, q) &=& e(k + q) a(t, k - q) + e^2 b(t, k - q) + ies \varepsilon(t, k - q), \label{eq:dirac_int_L}
\end{eqnarray}
and $a (t, k)$, $b (t, k)$, and $\varepsilon (t, k)$ are the Fourier transforms of $A_x (t,x)$, $A_x^2 (t,x)$, and $E (t,x)$, respectively. In what follows we assume that all the functions relate to the same value of $\boldsymbol{p}_\perp$ and use the identity $G({}_\zeta|{}^\kappa)_{\boldsymbol{p}, r;\, \boldsymbol{q}, r^\prime} = \delta (\boldsymbol{p}_\perp - \boldsymbol{q}_\perp) \, \delta_{r, r^\prime} g({}_\zeta|{}^\kappa)_{p_x, q_x, r}$. According to Eq.~(\ref{eq:G_inner_product_2}), the function ${}^+ \Psi_{p_x, r}$ can be represented as
\begin{equation}
{}^+ \Psi_{p_x, r} (t, \boldsymbol{x}) = \int \limits_{-\infty}^\infty \! \mathrm{d}k \, {}_- \Psi_{k, r} (t, \boldsymbol{x}) \, g({}_-|{}^+)_{k, p_x, r} + \int \limits_{-\infty}^\infty \! \mathrm{d}k \, {}_+ \Psi_{k, r} (t, \boldsymbol{x}) \, g({}_+|{}^+)_{k, p_x, r}.
\label{eq:Psi_plus}
\end{equation}
This yields
\begin{equation}
{}^+ \varphi_{p_x} (t, x) = \int \limits_{-\infty}^\infty \! \mathrm{d}k \, {}_- \varphi_{k} (t, x)\, g({}_-|{}^+)_{k, p_x, r} + \int \limits_{-\infty}^\infty \! \mathrm{d}k \, {}_+ \varphi_{k} (t, x)\, g({}_+|{}^+)_{k, p_x, r}.
\label{eq:phi_plus_decomp}
\end{equation}
Using the relations~(\ref{eq:phi_in}) and~(\ref{eq:fourier_out}) one can obtain for $t \leq t_\text{in}$
\begin{equation}
{}^+ f_{p_x} (t, k) = {}_- C_{k} \mathrm{e}^{ip_0 (k) (t - t_\text{in})} g({}_-|{}^+)_{k, p_x, r} + {}_+ C_{-k} \mathrm{e}^{-ip_0 (k) (t - t_\text{in})} g({}_+|{}^+)_{-k, p_x, r},
\label{eq:f_plus}
\end{equation}
where $p_0 (k) = \sqrt{k^2 + \pi_\perp^2}$. In order to find the matrix $g({}_\zeta|{}^+)$ we should propagate the function ${}^+ f_{p_x}$ backwards in time and for each value of $k$ decompose it into the two terms on the right-hand side [the normalization coefficients are given by Eq.~(\ref{eq:C_in_out})]. Once the solution for a given value of $p_x$ is found, it is easy to get a whole column $(k,~p_x)$ of the matrix $g({}_\zeta|{}^+)$. This is an important advantage of the method based on the Fourier transformation. The number of electrons created as a function of their momentum $p_x$ (with a given value of $\pi_\perp$ and $r$) can be evaluated as
\begin{equation}
\frac{(2\pi)^2}{V_\perp} \, n^-_{p_x, r} = \int \limits_{-\infty}^\infty \! \mathrm{d}k \, |g({}_-|{}^+)_{k, p_x, r}|^2.
\label{eq:n_g}
\end{equation}
In the following the results obtained for different field configurations will be presented.
\subsection{``Rectangular'' static field}\label{subsec:two_dim_rect_static}
First, we consider the ``rectangular'' static field given by
\begin{eqnarray}
E(t,x) &=& E \, \theta (T/2 - |t|) \, \theta (L - |x|), \label{eq:LT-constant_field}\\
A_x (t, x) &=& A_x (t)\, \theta (L - |x|), \label{eq:LT-constant_pot_tx}\\
A_x (t) &=& -E \Delta_T (t),
\label{eq:LT-constant_pot}
\end{eqnarray}
where the function
\begin{equation}
 \Delta_T (t) = 
\begin{dcases}
0 &\text{if}~~t \leq -T/2 = t_\text{in},\\
t - t_\text{in} = t + T/2 &\text{if}~~t_\text{in} = -T/2 \leq t \leq T/2 = t_\text{out},\\
t_\text{out} - t_\text{in} = T &\text{if}~~t \geq T/2 = t_\text{out},
\end{dcases}
\label{eq:delta_T}
\end{equation}
is introduced for convenience. In this case the following Fourier transforms are obtained:
\begin{eqnarray}
\varepsilon (t, k) &=& \varepsilon(k) \, \theta (T/2 - |t|), \label{eq:LT-constant_eps}\\
a (t, k) &=& -\varepsilon(k) \Delta_T (t), \label{eq:LT-constant_a} \\
b (t, k) &=&  E\, \varepsilon(k) \Delta_T^2 (t), \label{eq:LT-constant_b}
\end{eqnarray}
where
\begin{equation}
\varepsilon(k) = \frac{E}{\pi} \, \frac{\sin(kL)}{k}.
\label{eq:eps_k_1}
\end{equation}
Accordingly, for $t_\text{in} \leq t \leq t_\text{out}$ we have
\begin{equation}
L (t, k, q) = \frac{eE}{\pi} \, \frac{\sin (k - q)L}{k - q} \big [ -(k + q)(t - t_\text{in}) + eE(t - t_\text{in})^2 + is \big ].
\label{eq:LT-constant_L}
\end{equation}
For $t \in [t_\text{in}, t_\text{out}]$ the field strength $E(t, x) = E(x)$ is static. Assuming that the function $E(x)$ of $x$ is symmetric for $|x| \leq L$ and equal to zero for $|x| \geq L$, the out solutions for $t \geq t_\text{out}$ simplify further,
\begin{equation*}
{}^\pm f_{p_x} (t, k) = {}^\pm C_{p_x} \mathrm{e}^{\mp ip_0 (t - t_\text{out})} \bigg [ \frac{1}{\pi} \int \limits_0^L \mathrm{d}x \, \cos \bigg [ (k \pm p_x)x - eT \int \limits_0^x \! E(x')\mathrm{d}x' \bigg ]
\end{equation*}
\begin{equation}
{}+ \bigg ( \delta (k \pm p_x) - \frac{1}{\pi} \, \frac{\sin(k \pm p_x)L}{k \pm p_x} \bigg ) \cos(eETL)  + \frac{1}{\pi} \sin (eETL) \cos(k \pm p_x)L \, \mathcal{P} \frac{1}{k \pm p_x} \bigg ],
\label{eq:f_symm}
\end{equation}
where $\displaystyle \mathcal{P} \frac{1}{k}$ stands for the principal-value integral $\displaystyle \mathcal{P} \frac{1}{k} \, (h) = \text{p.v.} \int \limits_\mathbb{R} \mathrm{d}k \, \frac{1}{k} \, h(k)$. For the ``rectangular'' case we finally obtain
\begin{equation*}
{}^\pm f_{p_x} (t, k) = {}^\pm C_{p_x} \mathrm{e}^{\mp ip_0 (t - t_\text{out})} \bigg [ \frac{1}{\pi} \, \frac{\sin(k \pm p_x - eET)L}{k \pm p_x - eET}
\end{equation*}
\begin{equation}
{}+ \bigg ( \delta (k \pm p_x) - \frac{1}{\pi} \, \frac{\sin(k \pm p_x)L}{k \pm p_x} \bigg ) \cos(eETL)  + \frac{1}{\pi} \sin (eETL) \cos(k \pm p_x)L \, \mathcal{P} \frac{1}{k \pm p_x} \bigg ],
\label{eq:LT-constant_f_in}
\end{equation}
The functions ${}^\pm f_{p_x} (t, k)$ are generalized functions (distributions). If $L$ tends to infinity, the first term in square brackets tends to $\delta (k \pm p_x - eET)$ while the other terms disappear (for the last term one should employ the Riemann-Lebesgue lemma). In order to treat these functions properly for the numerical computation, one can first approximate the delta function by a ``triangular'' function with a very small support, and then choose the values of the parameters $E$, $T$, and $L$ so that the last term vanishes. This will allow one to treat the functions ${}^\pm f_{p_x} (t, k)$ as ordinary functions. However, it is more suitable to represent them as follows
\begin{equation*}
{}^\pm f_{p_x} (t, k) = {}^\pm C_{p_x} \mathrm{e}^{\mp ip_0 (t - t_\text{out})} \bigg [ \frac{1}{\pi} \, \frac{\sin(k \pm p_x - eET)L}{k \pm p_x - eET}
\end{equation*}
\begin{equation}
{}-\frac{1}{\pi} \, \frac{\sin(k \pm p_x - eET)L}{k \pm p_x + i \varepsilon} + \mathrm{e}^{ieETL} \delta (k \pm p_x) \bigg ],
\label{eq:LT-constant_f_in_2}
\end{equation}
where $\varepsilon \to 0$. This expression allows one to consider arbitrary values of $E$, $T$, and $L$.

\begin{figure}[h]
\center{\includegraphics[width=0.8\linewidth]{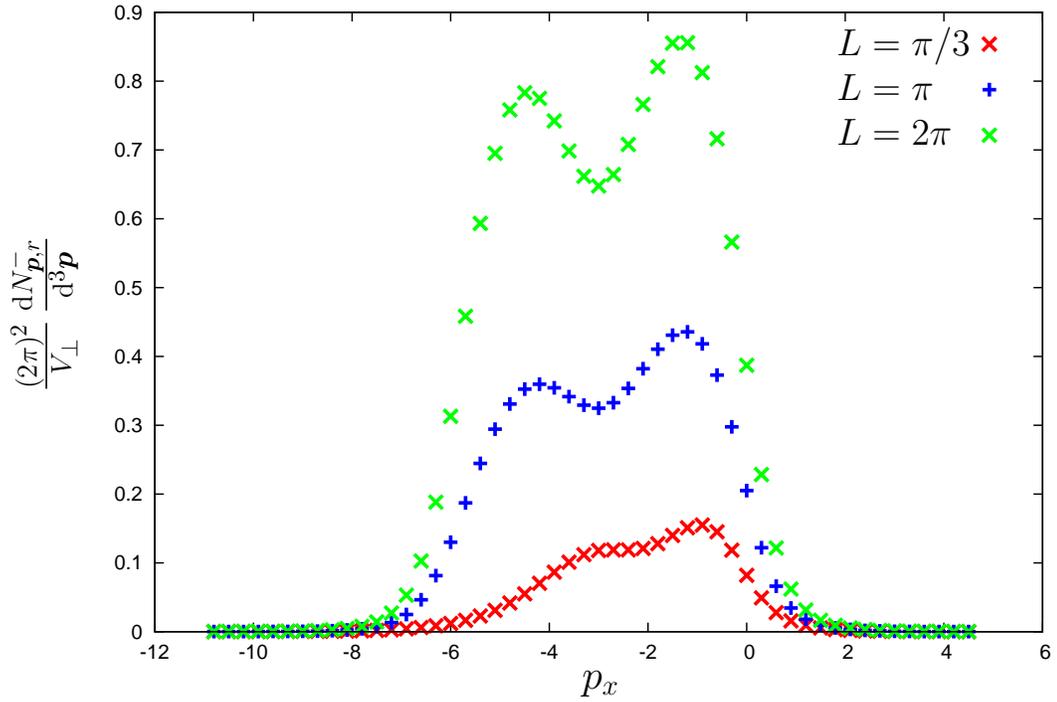}}
\caption{The mean number of electrons created by the ``rectangular'' static field ($T = 2.0$, $|e|E = 3.0$, $\pi_\perp = 1.0$) as a function of the momentum component $p_x$ for different values of $L$.}
\label{fig:rect_2dim_2}
\end{figure}

\begin{figure}[h]
\center{\includegraphics[width=0.8\linewidth]{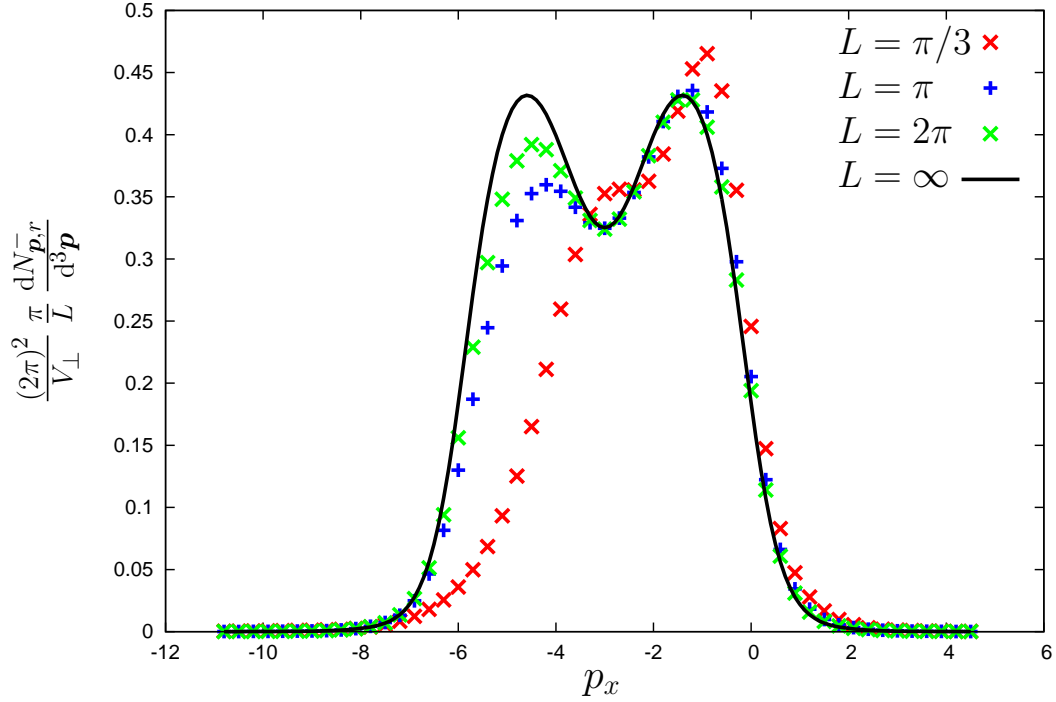}}
\caption{The mean number of electrons created by the ``rectangular'' static field ($T = 2.0$, $|e|E = 3.0$, $\pi_\perp = 1.0$), multiplied by the factor $\pi/L$, as a function of $p_x$ for different values of $L$. The numerical values tend to the analytical 1D ones (solid line) as $L$ increases.}
\label{fig:rect_2dim_2_norm}
\end{figure}

\indent In Fig.~\ref{fig:rect_2dim_2} the mean number of electrons created is displayed as a function of $p_x$ for $T = 2.0$, $|e|E = 3.0$, $\pi_\perp = 1.0$ and different values of $L$. It is obvious that these values being multiplied by $2\pi/2L = \pi/L$ tend to those obtained for the 1D case of a homogeneous external field and the corresponding pulse duration $T$, provided $L \to \infty$. These ``renormalized'' dependences are presented in Fig.~\ref{fig:rect_2dim_2_norm}.

\begin{figure}[h]
\center{\includegraphics[width=0.8\linewidth]{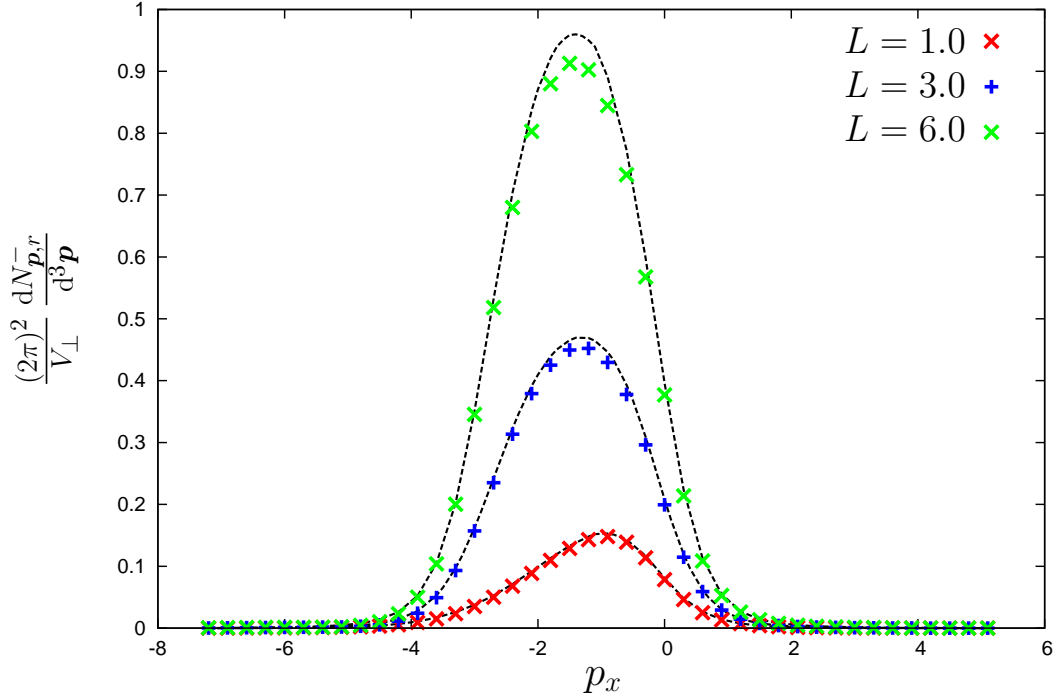}}
\caption{The mean number of electrons created by the ``rectangular'' static field ($T = 1.0$, $|e|E = 3.0$, $\pi_\perp = 1.0$) as a function of the momentum component $p_x$ for different values of $L$. The dashed lines relate to  $L = \pi/3$, $\pi$, and $2\pi$ while the crosses represent the spectra for $L = 1.0$, $3.0$, and $6.0$.}
\label{fig:rect_eps}
\end{figure}

\indent We employ the values $L = \pi/3$, $\pi$, and $2\pi$, since for the corresponding configurations the numerical convergence of the results with respect to the momentum grid step is achieved much faster, which allows us to minimize computational time. However, in order to demonstrate that the method can be employed for arbitrary values of the parameters $E$, $T$, and $L$, we compare the spectra for $L = \pi/3$, $\pi$, and $2\pi$ with those for $L = 1.0$, $3.0$, and $6.0$, respectively (see Fig.~\ref{fig:rect_eps}). We observe that the numerical results are stable with regard to $L$ and, therefore, arbitrary values of this parameter can be employed for further analysis.

\begin{figure}[h]
\center{\includegraphics[width=0.8\linewidth]{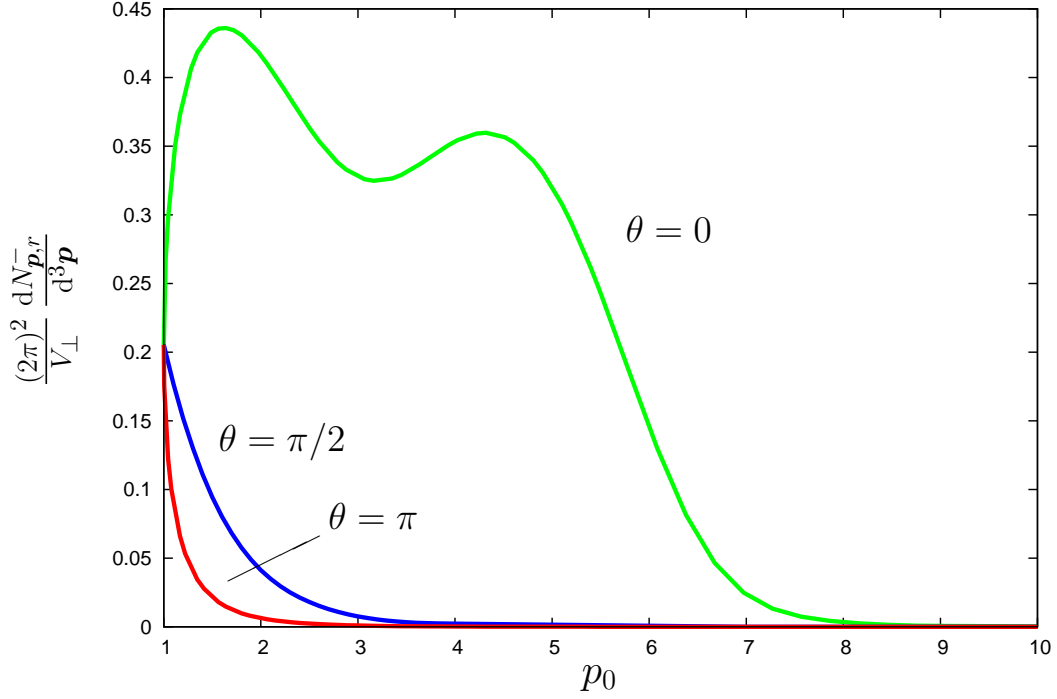}}
\caption{The mean number of electrons created by the ``rectangular'' static field ($T = 2.0$, $|e|E = 3.0$, $L = \pi$) as a function of their energy for different values of $\theta$.}
\label{fig:rect_2dim_2_ang}
\end{figure}

\indent Finally, we present the spectrum of the electrons as a function of their relativistic energy $p_0 = \sqrt{p_x^2 + \pi_\perp^2}$ for several values of the angle $\theta$ defined by $\cos \theta = p_x/|\boldsymbol{p}| = p_x/\sqrt{p_0^2 - m^2}$ (see Fig.~\ref{fig:rect_2dim_2_ang}). As was mentioned previously, the electrons created mostly have momenta along the negative direction of the $x$ axis ($\theta = 0$). As the angle $\theta$ becomes larger, the mean number of particles decreases: the number of electrons travelling perpendicularly to the electric field lines ($\theta = \pi/2$) is greater than the number of electrons moving along the field ($\theta = \pi$).
%
\subsection{``Rectangular'' time-dependent field}\label{subsec:two_dim_rect_t}
We now turn to the analysis of the pair creation for time-varying external electric field configurations being confined and uniform within the space region $-L \leq x \leq L$:
\begin{eqnarray}
E (t, x) &=& E(t) \, \theta (T/2 - |t|) \, \theta (L - |x|), \label{eq:L-constant_field}\\
A_x (t, x) &=& A_x (t) \, \theta (L - |x|), \label{eq:L-constant_pot_tx}\\
A_x (t) &=&
\begin{cases}
0 &\text{if}~~t \leq t_\text{in},\\
-\int \limits_{t_\text{in}}^{t} E(t') \mathrm{d} t' &\text{if}~~t_\text{in} \leq t \leq t_\text{out},\\
-\int \limits_{t_\text{in}}^{t_\text{out}} E(t') \mathrm{d}t' &\text{if}~~t \geq t_\text{out},
\end{cases} \label{eq:L-constant_pot}
\end{eqnarray}
where $t_\text{in} = -t_\text{out} = -T/2$. The Fourier transforms in Eq.~(\ref{eq:dirac_int_L}) take the form
\begin{eqnarray}
a (t, k) &=& A_x(t) \, \frac{1}{\pi} \, \frac{\sin(kL)}{k}, \label{eq:L-constant_a} \\
b (t, k) &=& A_x^2 (t) \, \frac{1}{\pi} \, \frac{\sin(kL)}{k}, \label{eq:L-constant_b} \\
\varepsilon (t, k) &=& \frac{E(t)}{\pi} \, \frac{\sin(kL)}{k} \, \theta (T/2 - |t|). \label{eq:L-constant_eps}
\end{eqnarray}
Thus, the function $L(t, k, q)$ for $t_\text{in} \leq t \leq t_\text{out}$ appears as
\begin{equation}
L (t, k, q) = \frac{e}{\pi} \, \frac{\sin (k - q)L}{k - q} \big [ (k + q)A_x (t) + eA_x^2 (t) + isE(t) \big ].
\label{eq:L-constant_L}
\end{equation}
The out solutions for $t \geq t_\text{out}$ can be obtained from Eq.~(\ref{eq:LT-constant_f_in}) by substituting $-ET \to A_x (t_\text{out})$.

\begin{figure}[h]
\center{\includegraphics[width=0.8\linewidth]{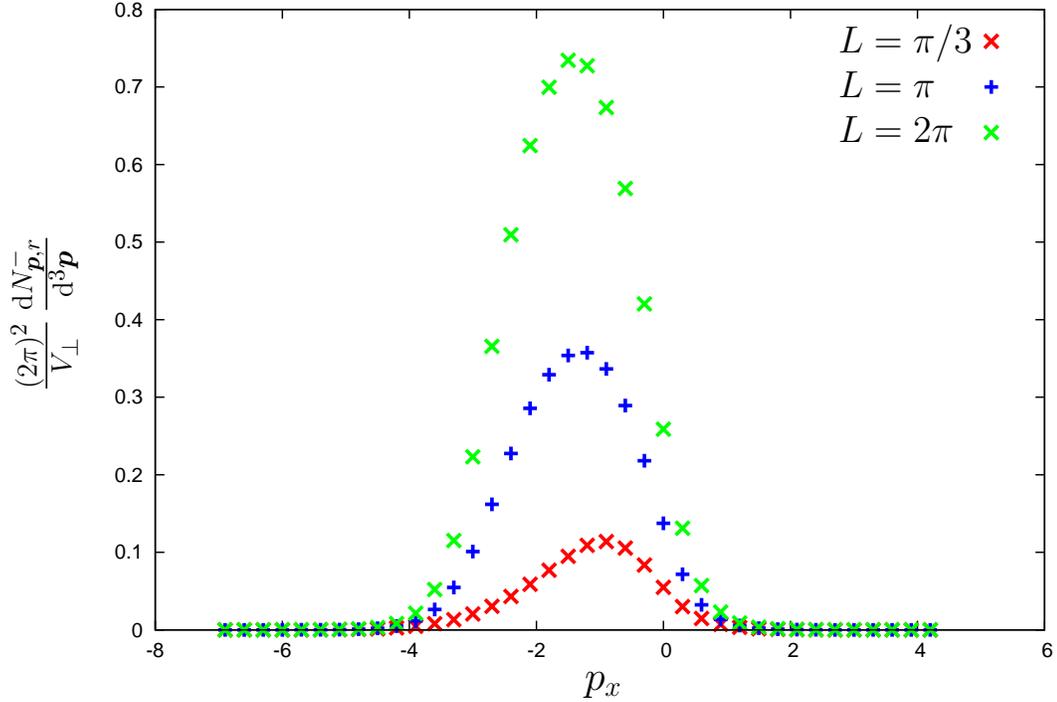}}
\caption{The mean number of electrons created by the ``rectangular'' Sauter-like field ($|e|E = 3.0$, $\tau = 0.5$, $\pi_\perp = 1.0$) as a function of their momentum $x$ component for different values of $L$.}
\label{fig:rect_sauter}
\end{figure}

\begin{figure}[h]
\center{\includegraphics[width=0.8\linewidth]{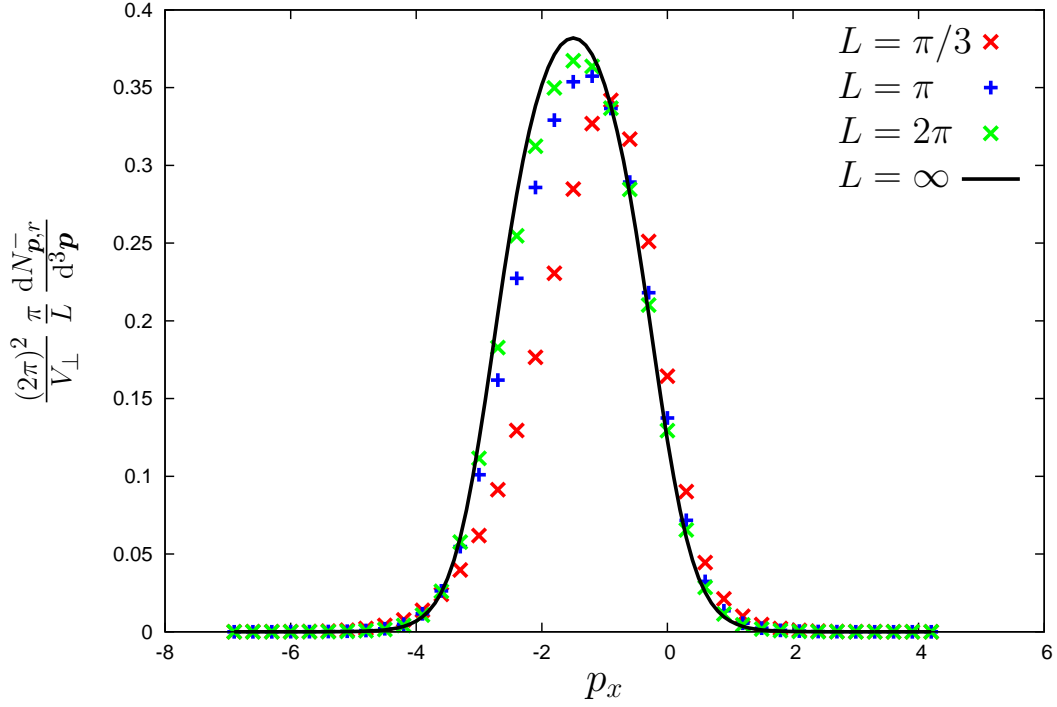}}
\caption{The mean number of electrons created by the ``rectangular'' Sauter-like field ($|e|E = 3.0$, $\tau = 0.5$, $\pi_\perp = 1.0$), multiplied by the factor $\pi/L$, as a function of $p_x$ for different values of $L$. The solid line represents the analytical 1D result for $L = \infty$.}
\label{fig:rect_sauter_norm}
\end{figure}

\indent As an example, we consider the Sauter-like potential depending on a parameter $\tau$ and for which $t_\text{in/out} \to \mp \infty$ (and $T \to \infty$):
\begin{eqnarray}
E(t) &=& \frac{E}{\cosh^2 (t/\tau)}, \label{eq:L-constant_Sauter_E}\\
A_x(t) &=& -E\tau \big ( 1 + \tanh (t/\tau) \big ). \label{eq:L-constant_Sauter_A}
\end{eqnarray}
The spectra of electrons created are depicted in Fig.~\ref{fig:rect_sauter} for $|e|E = 3.0$, $\tau = 0.5$, $\pi_\perp = 1.0$ and different values of $L$. In Fig.~\ref{fig:rect_sauter_norm} we present the spectra normalized by the factor $\pi/L$ and compare them with the analytical 1D result.

\begin{figure}[h]
\center{\includegraphics[width=0.8\linewidth]{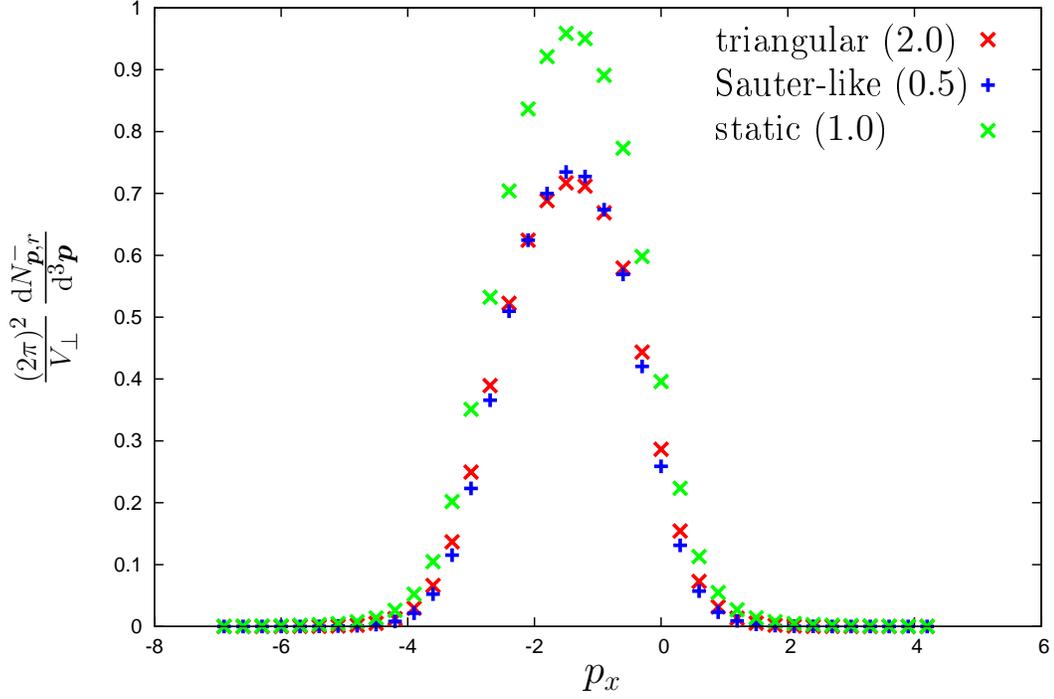}}
\caption{The mean number of electrons created by the Sauter-like field with $\tau = 0.5$ (blue points), the static field with $T = 1.0$ (green points), and the ``triangular'' field with $T = 2.0$ (red points). For all these dependences $|e|E = 3.0$, $\pi_\perp = 1.0$, and the field is assumed to be uniform inside a capacitor with $L = 2\pi$.}
\label{fig:rect_sauter_comparison}
\end{figure}

\indent Next, we examine a ``triangular'' field configuration:
\begin{equation}
E(t) =
\begin{cases}
E(1 + 2t/T) &\text{if}~~-T/2 \leq t < 0, \\
E (1 - 2t/T) &\text{if}~~0 \leq t \leq T/2.
\end{cases}
\label{eq:t_triang}
\end{equation}
In Fig.~\ref{fig:rect_sauter_comparison} we compare the spectra for the case of the static field with $T = 1.0$, the Sauter-like field with $\tau = 0.5$, and the ``triangular'' field configuration with $T = 2.0$. Since the integral $\int \limits_\mathbb{R} E(t) \mathrm{d} t$ and parameter $L$ have the same values for all these configurations, this comparison allows one to investigate what effect the temporal shape of the field strength has on the electronic spectrum. It turns out that the number of particles created is almost independent of the field strength shape, provided it is a continuous function of time, while in the case of a sharp ``rectangular'' pulse the number of particles is much greater. These results confirm that the switch-on and -off effects may play a very important role (see also Refs.~\cite{adorno_phys_scr_2015, adorno_2015, adorno_2016, linder_prd_2015}). Such effects with respect to the spatial finiteness of external fields will be discussed below.

\subsection{``Triangular'' static field}\label{subsec:two_dim_triangular}
It is also possible to examine external fields which have different spatial dependences. As an example, we analyze the ``triangular'' static field defined as
\begin{eqnarray}
E (t, x) &=& E(x) \, \theta (T/2 - |t|) \, \theta (L - |x|), \label{eq:E_triang}\\
E(x) &=&
\begin{cases}
E(1 + x/L) &\text{if}~~-L \leq x < 0, \\
E (1 - x/L) &\text{if}~~0 \leq x \leq L.
\end{cases} \label{eq:E_triang_2}
\end{eqnarray}
In this case
\begin{eqnarray}
\varepsilon (t, k) &=& \frac{E}{\pi L} \, \frac{1 - \cos(kL)}{k^2} \, \theta (T/2 - |t|), \label{eq:triang_eps}\\
a (t, k) &=& - \frac{E}{\pi L} \, \frac{1 - \cos(kL)}{k^2} \Delta_T (t), \label{eq:triang_a} \\
b (t, k) &=&
\begin{dcases}
0 &\text{if}~~t \leq t_\text{in},\\
\frac{2E^2}{\pi L} (t - t_\text{in})^2\, \frac{1}{k^2} \Big (1 - \frac{\sin(kL)}{kL} \Big ) &\text{if}~~t_\text{in} \leq t \leq t_\text{out},\\
\frac{2E^2}{\pi L} (t_\text{out} - t_\text{in})^2\, \frac{1}{k^2} \Big (1 - \frac{\sin(kL)}{kL} \Big )  &\text{if}~~t \geq t_\text{out},
\end{dcases} \label{eq:triang_b}
\end{eqnarray}
where $t_\text{in} = -t_\text{out} = -T/2$. Note that because the function $E(x)$ is now continuous, the Fourier transforms decrease faster, when $k \to \pm\infty$. For $t_\text{in} \leq t \leq t_\text{out}$ the function $L(t, k, q)$ reads
\begin{equation}
L (t, k, q) = \frac{eE}{\pi L} \, \frac{1}{(k - q)^2} \bigg [ \big (-(k + q)(t - t_\text{in}) + is \big ) \big (1 - \cos (k - q)L \big ) + 2eE(t - t_\text{in})^2 \bigg ( 1 - \frac{\sin (k - q) L}{(k - q)L}\bigg )\bigg ].
\label{eq:triang_L}
\end{equation}
The ``diagonal'' values are given by
\begin{equation}
L (t, k, k) = \frac{eEL}{\pi} \bigg [ \frac{1}{2} \big (-2k(t - t_\text{in}) + is \big ) + \frac{1}{3}eE(t - t_\text{in})^2 \bigg ].
\label{eq:L_diag}
\end{equation}
The out solutions have the following asymptotic behavior:
\begin{equation*}
{}^\pm f_{p_x} (t, k) = {}^\pm C_{p_x} \mathrm{e}^{\mp ip_0 (t - t_\text{out})} \bigg [ \frac{1}{\pi} \int \limits_0^L \mathrm{d}x \, \cos \Big [ (k \pm p_x - 2k_0)x + \frac{k_0}{L}x^2 \Big ]
\end{equation*}
\begin{equation}
{}+ \bigg ( \delta (k \pm p_x) - \frac{1}{\pi} \, \frac{\sin(k \pm p_x)L}{k \pm p_x} \bigg ) \cos(k_0 L)  + \frac{1}{\pi} \sin (k_0L) \cos(k \pm p_x)L \, \mathcal{P} \frac{1}{k \pm p_x} \bigg ],
\label{eq:triang_f_in}
\end{equation}
where $k_0 = eET/2$. The first (ordinary function) term in the external square brackets can be treated numerically.

\indent The results for the ``triangular'' electric field are displayed in Fig.~\ref{fig:triang} where the spectrum of electrons created for $T = 1.0$, $|e|E = 3.0$, and $L = 2\pi$ is compared with those for the ``rectangular'' static field and Sauter-like static field discussed below.
\subsection{Sauter-like static field}\label{subsec:two_dim_x_sauter}
In this subsection we investigate field configurations with a smooth Sauter-like spatial dependence:
\begin{equation}
E(x) = \frac{E}{\cosh^2 (x/\alpha)}.
\label{eq:E_x_sauter}
\end{equation}
Although it can be treated numerically as a function with a compact support (in this case Eq.~(\ref{eq:f_symm}) is valid), we can also evaluate all the necessary Fourier transforms analytically. For instance, by means of the residue theorem one can obtain
\begin{eqnarray}
\varepsilon (t, k) &=& \frac{E\alpha^2}{2}\, \frac{k}{\sinh(\alpha \pi k/2)} \, \theta (T/2 - |t|), \label{eq:x_sauter_eps}\\
a (t, k) &=& - \frac{E\alpha^2}{2}\, \frac{k}{\sinh(\alpha \pi k/2)} \Delta_T (t), \label{eq:x_sauter_a} \\
b (t, k) &=& \frac{E^2 \alpha^2}{12} \, \frac{k (4 + \alpha^2 k^2)}{\sinh(\alpha \pi k/2)} \, \Delta_T^2 (t). \label{eq:x_sauter_b}
\end{eqnarray}
The function $L(t, k, q)$ for $t_\text{in} \leq t \leq t_\text{out}$ takes the following form:
\begin{equation}
L (t, k, q) = \frac{eE\alpha^2}{2} \, \frac{k - q}{\sinh \big [ \alpha \pi (k-q)/2 \big ]} \Big [ - (k + q) (t - t_\text{in}) + \frac{eE}{6} \Big ( 4 + \alpha^2 (k-q)^2 \Big )(t - t_\text{in})^2 + is \Big ].
\label{eq:x_sauter_L}
\end{equation}
Since the function $E(x)$ is smooth, all the Fourier transforms as functions of $k$ decrease faster than any power function when $k \to \pm \infty$. The out solutions should be determined for $t \geq t_\text{out}$ using a regularization:
\begin{eqnarray}
{}^\pm f^\varepsilon_{p_x} (t, k) &=& {}^\pm C_{p_x} \mathrm{e}^{\mp ip_0 (t - t_\text{out})} F_\varepsilon (k \pm p_x), \label{eq:x_sauter_f_gen}\\
F_\varepsilon (k) &=& \int\limits_{-\infty}^{\infty} \frac{\mathrm{d}x}{2\pi} \, \mathrm{e}^{ikx} \mathrm{e}^{-\varepsilon|x|} \mathrm{exp} \bigg [ - ieT \int \limits_0^x \! E(x')\mathrm{d}x' \bigg ]. \label{eq:x_sauter_F_def}
\end{eqnarray}
The integral can be expressed as follows:
\begin{equation}
F_\varepsilon (k) = \frac{\alpha}{4\pi}\, \mathrm{e}^{-iB} \Gamma \Big ( \frac{ik\alpha}{2} + \frac{\varepsilon \alpha}{2} \Big ) \Gamma \Big ( -\frac{ik\alpha}{2} + \frac{\varepsilon \alpha}{2} \Big ) \boldsymbol{\mathrm{M}} \Big (\frac{ik\alpha}{2} + \frac{\varepsilon \alpha}{2}, \varepsilon \alpha, 2iB \Big ),
\label{eq:x_sauter_F_M}
\end{equation}
where $B = -eET\alpha$ and
\begin{equation}
\boldsymbol{\mathrm{M}} (a, b, z) = \sum_{n=0}^\infty \frac{(a)_n \, z^n}{\Gamma (n+b)\,n!},~~\text{where}~~(a)_n = a (a+1)\dots (a+n-1),
\label{eq:x_sauter_M}
\end{equation}
is the Kummer's function which is entire in $a$, $b$, and $z$. Thus, $F_\varepsilon (k)$ is a regular function at any point $k\in \mathbb{R}\setminus \{0\}$ for any sufficiently small value of $\varepsilon$ (including $\varepsilon = 0$). The analysis of its asymptotic expansion for $k \to 0$ and $\varepsilon \to 0$ reveals the following behavior of the function~(\ref{eq:x_sauter_F_M}):
\begin{eqnarray}
&&F_\varepsilon (k) = S_\varepsilon (k) + R_\varepsilon (k), \label{eq:x_sauter_FSR}\\
&&S_\varepsilon (k) = \frac{1}{\pi} \, \frac{\varepsilon}{k^2 + \varepsilon^2}\cos B - \frac{1}{\pi} \, \frac{k}{k^2 + \varepsilon^2}\sin B, \label{eq:x_sauter_S_def}\\
&&\lim_{\varepsilon \to 0} \int \limits_{-\infty}^{\infty} \! R_\varepsilon (k) h(k) \mathrm{d}k = \int \limits_{-\infty}^{\infty} \! R_0 (k) h(k) \mathrm{d}k,
\label{eq:x_sauter_R}
\end{eqnarray}
where $h(k)$ is an arbitrary test function (e.~g., a smooth function with a compact support). This means that the generalized function $F(k)$ has the following form:
\begin{equation}
F(k) = \delta (k) \cos(B)  - \frac{1}{\pi} \, \mathcal{P} \frac{1}{k} \, \sin (B) + \text{ordinary function}.
\label{eq:x_sauter_F_form}
\end{equation}
As was expected, the ``generalized part'' of the out function is similar to that expressed by Eq.~(\ref{eq:f_symm}).

\begin{figure}[h]
\center{\includegraphics[width=0.8\linewidth]{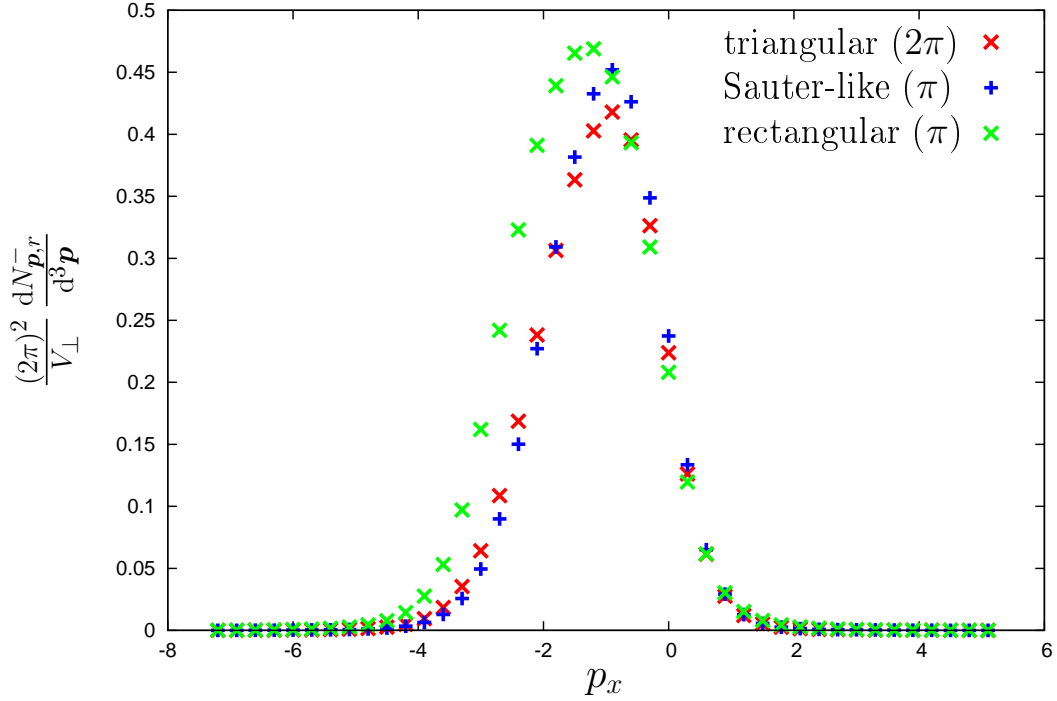}}
\caption{The mean number of electrons created by the ``triangular'' static field with $L = 2\pi$ (blue points), the ``rectangular'' static field with $L = \pi$ (green points), and the Sauter-like static field with $\alpha = \pi$ (red points). For all the spectra $|e|E = 3.0$, $T = 1.0$, and $\pi_\perp = 1.0$.}
\label{fig:triang}
\end{figure}

\indent In Fig.~\ref{fig:triang} the spectrum of electrons created is displayed for three different space-dependent field configurations: the ``rectangular'', ``triangular'', and Sauter-like fields. For these configurations the ``electrical work'' $\int \limits_{\mathbb{R}} E(x) \mathrm{d}x$ and the field duration $T$ are the same, so the figure allows one to analyze the spectrum with respect to different shapes of the field strength $E(x)$. One observes that the number of electrons created by the ``rectangular'' static field is considerably larger than that for ``continuous'' field configurations. This also indicates the importance of the switch-on and -off processes and demonstrates that the electronic spectrum strongly depends on whether the field-strength function $E(t, x)$ is continuous.
\subsection{Combination of static and oscillating fields in the $(1+1)$ case}\label{subsec:two_dim_static_oscillations}
If one neglects the transverse degrees of freedom, the total number $N$ of electrons (pairs) created can also be found by means of the procedure described above using $\pi_\perp = m$. In this case the quantum number $r$ disappears and instead of Eq.~(\ref{eq:n_g}) one has to use
\begin{equation}
n_{p_x} = \int \limits_{-\infty}^\infty \! \mathrm{d}k \, |g({}_-|{}^+)_{k, p_x, r}|^2, \quad N = \int \limits_{-\infty}^\infty \! \mathrm{d}p_x \, n_{p_x}.
\label{eq:n_g_1}
\end{equation}
We will examine the configuration
\begin{equation}
E(t, x) = \frac{1}{\cosh^2 (x/\alpha)} \, \Big ( E_1 \sin \omega t + E_2 \Big ) \, \theta(T/2 - |t - T/2|),
\label{eq:E_jiang}
\end{equation}
which was proposed in Ref.~\cite{jiang_pra_2012}, where it was found that the total pair-creation rate can be significantly increased in comparison to that for the case when we have only one (either static or oscillating) pulse.

\begin{figure}[h]
\center{\includegraphics[width=0.8\linewidth]{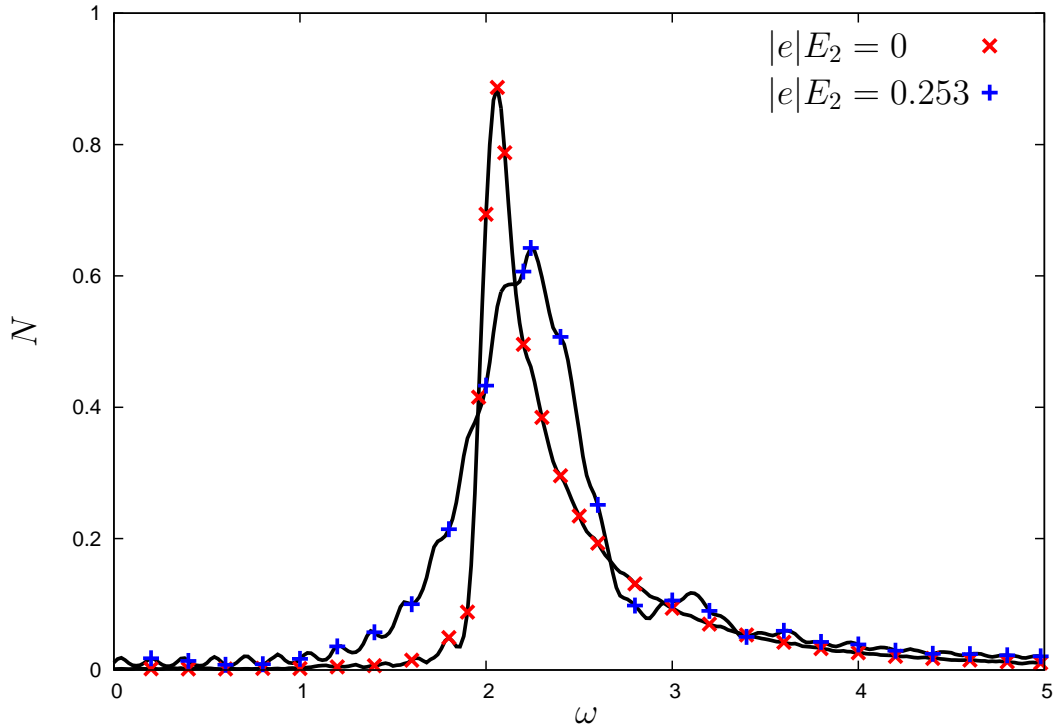}}
\caption{The total number of pairs created as a function of $\omega$ for $T = 37.56 = 0.002~\text{[au]}$, $\alpha = 5.0$, $|e|E_1 = 0.147$, and two different values of $E_2$ [see Eq.~(\ref{eq:E_jiang})]. The solid lines represent the corresponding results taken from Ref.~\cite{jiang_pra_2012}.}
\label{fig:jiang}
\end{figure}

\indent In Fig.~\ref{fig:jiang} the total number of pairs created is depicted as a function of the frequency $\omega$ for $T = 37.56 = 0.002~\text{[au]}$, $\alpha = 5.0$, $|e| E_1 = 0.147$, and two different values of $E_2$. The results are in a perfect agreement with the previous findings~\cite{jiang_pra_2012}, so our approach provides an independent verification for this nontrivial field configuration.
\subsection{Pair production in spatially periodic fields}\label{subsec:two_dim_periodic}
Finally, we will discuss the case of spatially periodic backgrounds for which our technique seems to be extremely efficient. Let $E(x)$ be the spatial dependence of the external field which obeys $E(x+d) = E(x)$ for all $x \in \mathbb{R}$ and given $d$. For simplicity we assume that this function is even. Therefore, it can be represented as
\begin{equation}
E(x) = \frac{a_0}{2} + \sum_{n=1}^\infty a_n \, \cos \bigg( \frac{2\pi}{d} n x\bigg).
\label{eq:E_per_fourier}
\end{equation}
Thus, its Fourier transform reads
\begin{equation}
\varepsilon (k) = \sum_{n=-\infty}^\infty \frac{a_n}{2} \, \delta \bigg (k + \frac{2\pi}{d} n \bigg),
\label{eq:epsilon_per}
\end{equation}
where $a_n = a_{-n}$ for negative $n$. It follows that the integral in Eq.~(\ref{eq:dirac_xt_phi_momentum}) can be analytically reduced to the discrete sum without any loss of accuracy:
\begin{equation}
\int \limits_{-\infty}^\infty \! \mathrm{d}q L(t, k, q) f_{p_x} (t, q) = \sum_{n=-\infty}^\infty c_n (t, k) f(t, k+2\pi n/d),
\label{eq:L_per}
\end{equation}
where the coefficients $c_n (t, k)$ can be easily obtained. The problem naturally becomes discrete and the computations can be carried out much faster, especially when the corresponding Fourier series converges rapidly (this fact was also pointed out in Ref.~\cite{woellert_2015}). Note, that the coefficients $a_n$ should be evaluated once in the very beginning, while solving Eq.~(\ref{eq:dirac_xt_phi_momentum}) now does not take much time.

\begin{figure}[h]
\center{\includegraphics[height=0.34\linewidth]{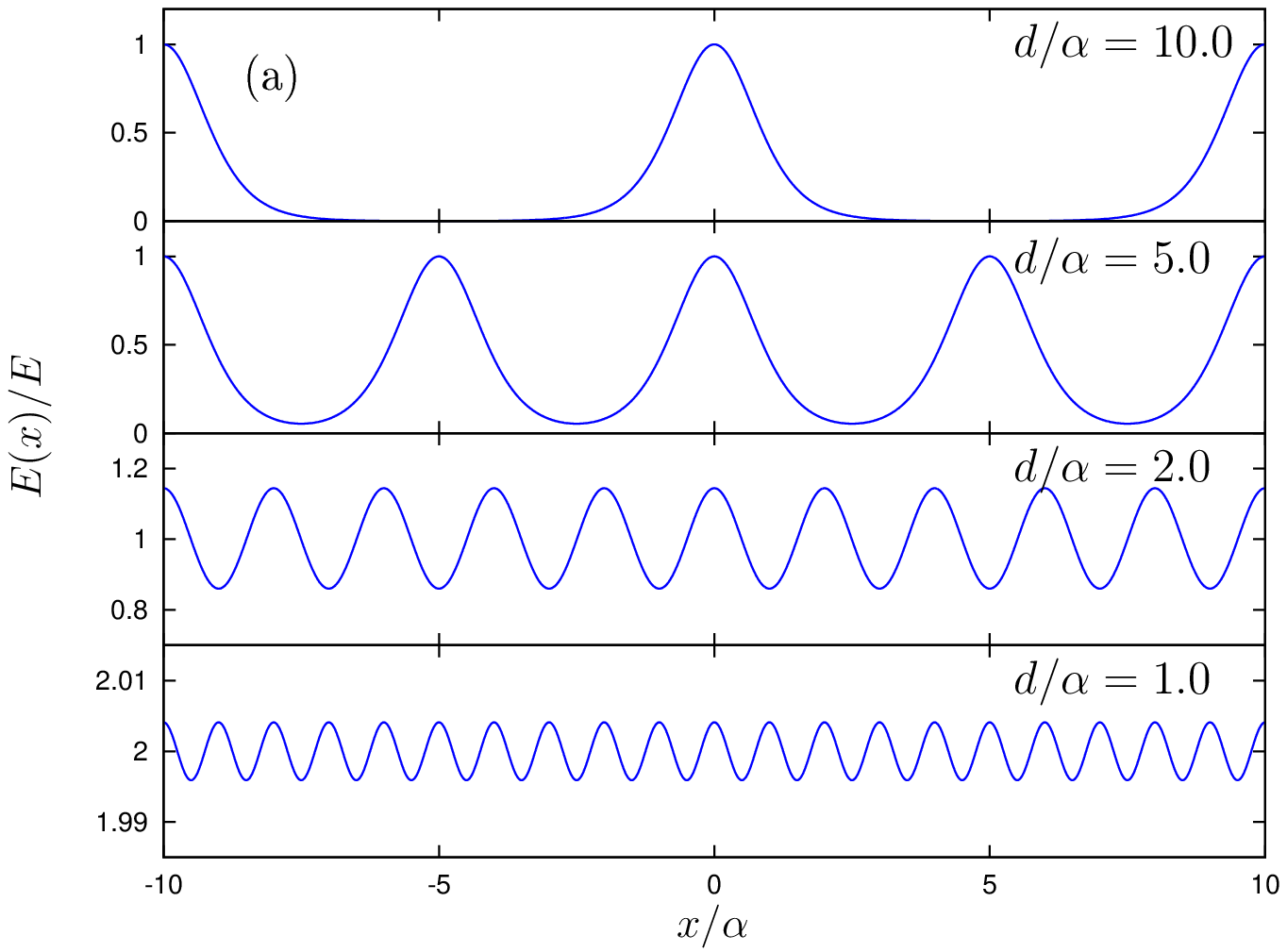}~~~
\includegraphics[height=0.34\linewidth]{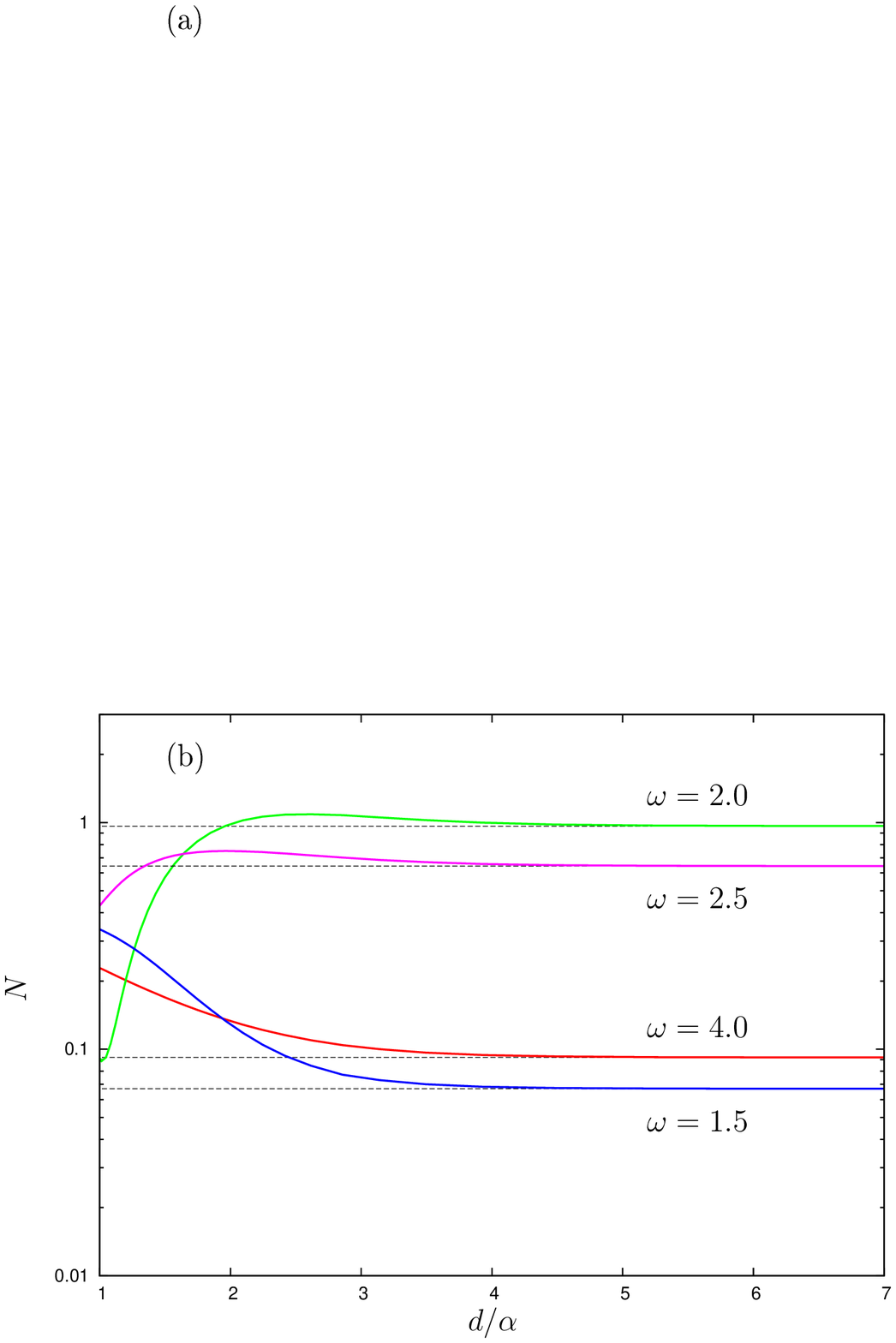}}
\caption{(a)~The electric field profile $E(x)$ given by Eq.~(\ref{eq:E_per_sauter}) for different values of $d$. (b)~The total number of pairs created per pulse as a function of $d$ for $\alpha = 4.0$, $T = 4 \pi$, and different values of $\omega$.}
\label{fig:per}
\end{figure}

\indent In order to provide an illustration, we will consider (again, in the $(1+1)$ case) an infinite sequence of the oscillating Sauter pulses:
\begin{equation}
E (t, x) = E(x) \, \theta(T/2 - |t - T/2|) \sin \omega t, \quad E(x) = \sum_{n=-\infty}^\infty \frac{E}{\cosh^2 \big [ (x + nd)/\alpha\big ] },
\label{eq:E_per_sauter}
\end{equation}
where $d$ is the distance between two nearest peaks. The function $E(x)$ for different ratios $d/\alpha$ is displayed in Fig.~\ref{fig:per}(a). The overall number $N$ of pairs created per pulse as a function of $d$ is presented in Fig.~\ref{fig:per}(b) for $\alpha = 4.0$, $T = 4 \pi$, and several values of $\omega$. As was expected, in the limit $d \to \infty$ the results tend to those for the case of a single pulse since pulses, separated by a very large distance, produce particles independently. If $d/\alpha$ becomes very small the pulses merge and the electric field strength growth unlimitedly. However, even when $d$ is comparable to $\alpha$ and the maximum field strength is almost the same as for a single pulse (e.~g., for $d/\alpha = 2$ $E_\text{max} \approx 1.14 E$), there is a considerable enhancement of the pair-creation rate for certain values of the field parameters.

\indent Finally, we note that the periodicity of the spatially localized external field can always be achieved artificially by multiplying the corresponding pulse. Using a large value of $d$, one can obtain all the necessary quantities for the case of an individual pulse.
%
\section{Production of scalar particles}
\label{sec:bose}
In the case of spinless particles the general formalism is quite similar to that described in Section~\ref{sec:theory} (see Refs.~\cite{fradkin_gitman_shvartsman, gmm_1994}). The in and out sets of solutions of the Klein-Fock-Gordon equation (i.~e., Eq.~(\ref{eq:dirac_squared}) without the last term) are orthogonal and complete with regard to the following inner product (in the temporal gauge $A_0 (x) = 0$):
\begin{equation}
(\Phi_1, \, \Phi_2) = i \int \Phi^*_1 (t, \boldsymbol{x}) (\overrightarrow{\partial_t} - \overleftarrow{\partial_t}) \Phi_2 (t, \boldsymbol{x}) \mathrm{d}^3\boldsymbol{x}
\label{eq:inner_product_bosons}
\end{equation}
Namely,
\begin{eqnarray}
&&({}_\zeta \Phi_n, \, {}_\kappa \Phi_m) = ({}^\zeta \Phi_n, \, {}^\kappa \Phi_m) = \zeta \delta_{\zeta \kappa} \delta_{nm},\quad \zeta, \kappa = \pm,  \label{eq:phi_orthonormality} \\
&&\sum_{n, \zeta} {}_\zeta \Phi_n (t, \boldsymbol{x}) \, {}_\zeta \Phi^*_n (t, \boldsymbol{x}') = \sum_{n, \zeta} {}^\zeta \Phi_n (t, \boldsymbol{x}) \, {}^\zeta \Phi^*_n (t, \boldsymbol{x}') = \delta (\boldsymbol{x} - \boldsymbol{x}'). \label{eq:phi_completeness}
\end{eqnarray}
The propagation function can be expanded in the following way:
\begin{equation}
G(x, x') = \sum_{\zeta,\, n} \zeta \, {}_\zeta \Phi_n (x) \, {}_\zeta \Phi^*_n (x') = \sum_{\zeta,\, n} \zeta \, {}^\zeta \Phi_n (x) \, {}^\zeta \Phi^*_n (x').
\label{eq:G_expansion_bose}
\end{equation}
Since Eqs.~(\ref{eq:G_inner_product_1}), (\ref{eq:G_inner_product_2}), (\ref{eq:num_el}), and (\ref{eq:num_pos}) remain the same, in order to analyze processes of scalar particle production we can immediately modify our method according to the following prescription~\cite{gav_git_prd_1996, adorno_2015}: one should set $s = 0$ and use the normalization factors given by
\begin{equation}
{}_\zeta C_{p_x} = {}^\zeta C_{p_x}  = (2\pi )^{-3/2} \, (2p_0)^{-1/2}
\label{eq:C_in_bose}
\end{equation}
instead of those displayed in Eq.~(\ref{eq:C_in_out}).

\begin{figure}[h]
\center{\includegraphics[width=0.95\linewidth]{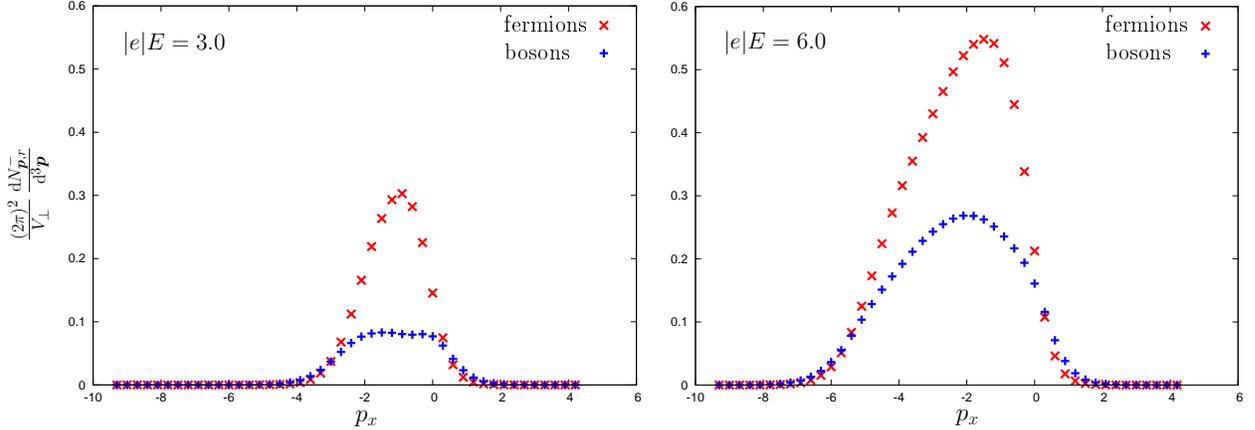}}
\caption{The mean number of negatively charged fermions (bosons) created by the space-time-dependent Sauter-like field~(\ref{eq:tx_sauter}) for $\tau = 0.5$, $\alpha = \pi$, $\pi_\perp = 1.0$ and two different values of $E$.}
\label{fig:bose_fermi}
\end{figure}

\indent In Fig.~\ref{fig:bose_fermi} we present the spectra of particles created by the space-time-dependent Sauter-like pulse
\begin{equation}
E(t,x) = \frac{E}{\cosh^2 (t/\tau)\cosh^2 (x/\alpha)}
\label{eq:tx_sauter}
\end{equation}
for $\tau = 0.5$, $\alpha = \pi$, $\pi_\perp = 1.0$ and two different values of $E$ for both the fermionic and bosonic cases. The support of the spectrum of negatively charged scalar particles coincides with that of electrons, while the total number of spinless particles is smaller. The latter property can be found even when the field configuration relates to the pure time-dependent Sauter field.
%
\section{Discussion and conclusion}
\label{sec:discussion}
In the present paper the phenomenon of electron-positron pair production in time- and space-inhomogeneous external electric fields was considered. A new efficient numerical technique, which can be used to obtain the corresponding spectra of particles created, was described in great detail. Our approach was applied to a number of various 1D and 2D backgrounds and the results were compared with exact analytical expressions. It was shown that the method can be used to analyze numerous 2D field configurations that do not have any exact solutions. We also examined different temporal and spatial shapes of the electric pulse. It was confirmed that the switch-on and -off effects may be very significant (see also Refs.~\cite{adorno_phys_scr_2015, adorno_2015, adorno_2016, linder_prd_2015}), especially if one considers a ``sharp'' time- or space-dependent switching function, which means that for 2D backgrounds both temporal and spatial dependences should be taken into account exactly which can be done with the aid of the technique developed. In this context it is worth noting that, according to Ref.~\cite{gies_prl_2016}, the onset of pair production in the presence of a static but spatially inhomogeneous field exhibits a scaling behavior near the critical point independently of the microscopic profile details.

\indent Besides, in the present paper it was shown that the approach is especially profitable when one considers spatially periodic backgrounds. This feature may be extremely helpful for the analysis of different laser field configurations (see Ref.~\cite{woellert_2015}). Finally, it was demonstrated that the analogous calculations can be easily carried out for the case of scalar QED.

\indent As was pointed out in Ref.~\cite{schutzhold_prl_2008}, the pair-production rate can be dramatically increased by superimposing a strong and slowly changing field by a weak and fast-varying pulse. In the recent paper \cite{schneider_2016} this dynamical assistance was investigated in the presence of a two-dimensional background. However, since the slowly varying pulse acts for a very long time, our numerical procedure becomes much more time consuming. The corresponding analysis of the dynamically assisted Schwinger effect based on the numerical approach presented here will be an important subject for future investigations.

\indent Another way to drastically increase the mean number of pairs was proposed in Ref.~\cite{dunne_prd_2009}, where the combination of a plane-wave x-ray beam and a focused optical laser pulse was considered. It may be possible to provide the corresponding experimental study at the Extreme Light Infrastructure (ELI) facility~\cite{eli}. In Ref.~\cite{dunne_prd_2009} it was shown that the presence of a hard x-ray photon with frequency $\omega \sim 2 m$ leads to an enhancement of order $\mathrm{exp} [2m^2/(|e|E)]$ (in the weak-field regime $|e|E \ll m^2$). This result was obtained by evaluating the imaginary part of the polarization tensor in a constant external electric field. Going beyond this approximation, e.~g., taking into account the temporal dependence of the laser field, should be very important for planning measurements. We expect that our numerical technique will be able to serve this purpose. Assuming the laser field to be classical and including the interaction between the quantized Dirac and electromagnetic fields, one can rewrite Eq.~(\ref{eq:num_el}) in the case of one initial photon as
\begin{equation}
n^-_m = \langle 0,\text{in} | \hat{c}_{\boldsymbol{k}, \lambda} S^\dagger \hat{b}^\dagger_m (\text{out}) \hat{b}_m (\text{out}) S \hat{c}^\dagger_{\boldsymbol{k}, \lambda} |0,\text{in}\rangle, \label{eq:num_S}
\end{equation}
where $S$ is the scattering matrix in the external field and $\hat{c}^\dagger_{\boldsymbol{k}, \lambda}$ is the photon creation operator. To zeroth order this leads to the expression~(\ref{eq:num_el}). Once we have the in and out sets of solutions and elements of the $G$ matrices, it is possible to evaluate Eq.~(\ref{eq:num_S}) to a given order in $\alpha$ (in fact, within the scenario from Ref.~\cite{dunne_prd_2009} the first-order contribution dominates). Furthermore, this approach requires the integrations over spatial coordinates which can be easily reduced to simple convolutions in the momentum representation. Although the corresponding calculations generally appear to be very tedious, at least in the 1D or 2D case they might become possible.
\section*{Acknowledgments}
This investigation was supported by RFBR (Grant No.~16-02-00334) and by Saint Petersburg State University (SPbU) (Grants No.~11.42.987.2016, 11.42.939.2016, 11.38.269.2014, and 11.38.237.2015). I. A. A. acknowledges the support from the German-Russian Interdisciplinary Science Center (G-RISC) funded by the German Federal Foreign Office via the German Academic Exchange Service (DAAD), from TU Dresden (DAAD-Programm Ostpartnerschaften), and from the ``Dynasty'' foundation.

\end{document}